# Angle-of-Attack Modulation in Trajectory Tracking for a Reusable Launch Vehicle

Ran Zhang[1]  Huifeng Li[2]  and Rui Zhang[3]

*Beihang University, Beijing, China, 100191*

**This paper deals with the problem of angle-of-attack modulation with the aim of enhancing transient performance of entry guidance during bank reversals, while compensating adverse effects of fast time-varying transient disturbances. An extended single-input/single-output system is developed in the velocity domain by means of a dynamic extension technique, and explicitly captures the trajectory dynamics of angle-of-attack modulation. A normal form for this extended system is derived for the sake of employing a feedback linearization controller. Further, the control characteristics of angle-of-attack modulation is found to be a non-minimum phase behavior under two common conditions in a near- equilibrium glide flight. Therefore, the issue of angle-of-attack modulation is formulated as robust output stabilization of the non-minimum phase system. A disturbance observer-based feedback linearization technique is used to design a robustly dynamical output-feedback controller for angle-of-attack modulation, and an internal-state feedback controller for bank-angle modulation is used to stabilize the unstable internal dynamics. Numerical simulations are conducted to demonstrate that the performance of the proposed method of angle-of-attack modulation is enhanced compared to the existing shuttle method.**

---

[1] Assistance Professor, School of Astronautics, zhangran@ buaa.edu.cn.

[2] Professor, School of Astronautics, lihuifeng@buaa.edu.cn.

[3] Assistance Professor, School of Astronautics, zhangrui@buaa.edu.cn.

## Nomenclature

| | |
|---|---|
| *A*, *B* | = matrices describing the linearized trajectory dynamics |
| $C_L$, $C_D$ | = lift and drag coefficients, respectively |
| *C* | = output matrix in the output equation $C = \begin{bmatrix} c_r & c_\gamma \end{bmatrix} = \begin{bmatrix} \hat{D}_r & 0 \end{bmatrix}$ |
| *D* | = nondimensional drag acceleration |
| $g_0$ | = gravitational acceleration |
| *H* | = atmosphere scale height |
| *L* | = nondimensional lift acceleration |
| *M* | = Mach number |
| *m* | = vehicle mass |
| $R_e$ | = radius of Earth |
| ***e*** | = tracking-error vector |
| *r* | = radial distance from Earth's center to the vehicle, normalized by $R_e$ |
| $S_{\text{ref}}$ | = aerodynamic reference area |
| *V* | = Earth-relative velocity, normalized by $\sqrt{g_0 R_e}$ |
| *α* | = angle-of-attack |
| *γ* | = flight-path angle |
| *ξ* | = damping ratio |
| *ρ* | = atmospheric density |
| $\rho_0$ | = sea-level atmospheric density |
| *σ* | = bank angle |
| *t* | = time |
| *τ* | = nondimensional time $t / \sqrt{R_e/g_0}$ |
| ***x*** | = state vector of a longitudinal trajectory in the velocity domain $\boldsymbol{x} = [r \quad \gamma]^T$ |
| ***u*** | = control input vector $[\alpha \quad \sigma]^T$ |
| *ϕ* | = latitude |
| *ψ* | = velocity azimuth angle, measured from the north in a clockwise direction |
| *Ω* | = self-rotation rate of Earth, normalized by $\sqrt{g_0/R_e}$ |
| $\omega_n$ | = natural frequency in the velocity domain |

$\delta u_1$ = angle-of-attack deviation, $\delta u_1 = u_1 - \hat{u}_1$

$\chi$ = feed-forward term in the output equation $\chi = \hat{D}_\alpha$

*Subscripts*

$r$ = differentiation with respect to $r$

$\alpha$ = differentiation with respect to $\alpha$

*Superscripts*

$'$ = differentiation with respect to $V$

$\hat{\ }$ = value of the reference trajectory

$-$ = elements of the simplified linear model

## I. Introduction

For a Reusable Launch Vehicle (RLV), the problem of tracking a reference trajectory and rejecting unwanted disturbances and uncertainties is one of the defining problems of entry guidance. In dealing with the trajectory tracking, feasible trajectory control means consist of bank-angle and angle-of-attack modulation [1, 2]: bank-angle modulation refers to the use of the bank angle magnitude to modulate the vertical component of the aerodynamic lift on the vehicle; angle-of-attack modulation refers to the use of the angle-of-attack deviation from its reference profile to adjust the aerodynamic coefficients. The choice of these two means is closely related to specific subsystems of the RLV, particularly involving the thermal protection system (TPS) and the attitude control system (ACS) [3]. Many of typical applications of the two control means can be found in the design of the shuttle entry guidance [1, 2] and X-33 entry guidance methods [4-8]. In practice, bank-angle modulation is prone to induce undesired large oscillations and results in a series of detrimental influences on other subsystems of the RLV [9, 10], due to a limited magnitude of the bank-angle-angular rate during bank reversals [11]; besides, bank-angle modulation cannot timely compensate adverse effects caused by fast time-varying transient disturbances, such as density gradients and drag acceleration segment transitions, because of its inherent long-period control characteristics. Capable of making up for these shortcomings, angle-of-attack modulation provides a supplementary control means in a short-period timescale, thereby enhancing both transient guidance performance (such as tracking accuracy and phugoid-oscillation attenuation) and robustness to the fast time-varying disturbances. In the literature, an intuitive method of angle-of-attack modulation is first proposed in the shuttle entry guidance method [1, 2] and then is successfully applied to many of entry guidance methods [4, 7]; in addition, an aerocapture guidance algorithm uses this method to improve flight performance, and the numerical simulations illustrate its benefits in reducing the aerocapture risk and the propellant consumption needed to reach the high energy target orbit [12]. Most of RLV entry guidance methods are based on a cooperative control strategy: bank-angle modulation is the primary control means and angle-of-attack modulation is the secondary control means that is used to compensate for the short-period drag acceleration profile deviation from the reference profile. To the best of the authors' knowledge, however, rigorous analysis and systematic design methodologies are not available for angle-of-attack modulation. The motivation of this paper is to formally formulate the problem of angle-of-attack modulation and then to develop a practical controller to solve it. A mathematically rigorous formulation will change the current sole reliance on the heuristic arguments for the use of angle-of-attack modulation and yield systematic guidelines for using more advanced control approaches to modulate the angle-of-attack, thus improving the capability of entry guidance.

In this paper, we focus primarily on formulating the problem of angle-of-attack modulation using the control theory of output regulation/stabilization, especially regarding a non-minimum phase control challenge, and on developing a disturbance observer-based controller to achieve angle-of-attack modulation. An extended single-input/single-output (SISO) system of which output is the drag acceleration tracking error and control input is the second-order derivative of the angle-of-attack deviation is first developed by a dynamic extension technique [12], and rigorously describes the trajectory dynamics of angle-of-attack modulation. Correspondingly, the issue of angle-of-attack modulation is formulated as a robust output-feedback stabilization problem, and a normal form for the extended SISO system is obtained that is composed of internal dynamics－the tracking-error dynamics－and external dynamics－the augmented dynamics. A feedback linearization method can effectively solve this kind of control problems, provided that the internal dynamics are stable [13-15]. In fact, two common conditions of a near-equilibrium glide flight rendering the internal dynamics unstable are identified by examining the stability of the resulting zero dynamics. Thus, the issue of angle-of-attack modulation is further formulated as the output-stabilization of the non-minimum phase system, which poses the main challenge to the feedback linearization method [14]. Specifically, the non-minimum phase property here implies that the actual angle-of-attack can easily diverge from its reference profile, if the internal dynamics in question are not actively stabilized; on the other hand, it is demanded that the actual angle-of-attack should be in a constrained neighborhood of its reference profile from the perspective of overall performance of the RLV [1-2]. In other words, the modulated angle-of-attack has the great risk to violate the constrained working range if the divergence cannot be suppressed. Thus, it is of great importance to actively stabilize the internal dynamics for the implementation of angle-of-attack modulation. Inspired by numerous control methodologies that have been proposed to address the stabilization problems for non-minimum phase systems [16-17], the work here adopts an internal-state feedback controller for bank-angle modulation to stabilize the internal dynamics; that is, utilize the bank angle to regulate the drag acceleration errors and then use the controlled drag acceleration errors to actively stabilize the internal dynamics. Such a stabilization methodology is a practical way to suppress the divergence of the actual angle-of-attack, and essentially the cooperative control strategy in the shuttle method can be regarded as a particular kind of this methodology. In addition, because modeling errors and disturbances cannot be explicitly considered in employing the feedback-linearization tool, the controller to be designed needs to deal with the robustness of the feedback-linearization [18]. So far, many of robust control approaches have been proposed to recover prescribed performance of the nominal feedback controllers; among these approaches the disturbance observer-based scheme is a practical solution [19]. Therefore, by the disturbance observer-based feedback linearization method, a robustly dynamical output-feedback controller for angle-of-attack modulation is devised to stabilize the external dynamics. At last, nominal and dispersion tests are performed to demonstrate the performance of the proposed method of angle-of-attack modulation.

The contributions of this paper are as follows. First, the trajectory dynamics of angle-of-attack modulation are taken into account using the extended SISO system. In comparison, the existing methods are based on the empirical justification that the trajectory dynamics have no effect on the drag acceleration during modulating the angle-of-attack, and thus the derivation directly uses the simplified static relationship between the drag coefficient and the angle-of-attack [1]. In fact, one serious problem caused by ignoring the trajectory dynamics have been encountered in numerical tests of the X-33 entry guidance design: the effect of angle-of-attack modulation can be opposite of what is desired [4]. Second, by examining the stability of the internal dynamics of the extended SISO system, the important control property is found that angle-of-attack modulation exhibits the non-minimum phase behavior under two common conditions in the near-equilibrium glide flight. Due to this undesired property, it is required that an effective control scheme should be devised to suppress the divergence of the modulated angle-of-attack. From the perspective of control theory, this property also gives a rigorous explanation of the function of the angle-of-attack deviation term in bank-angle modulation equation of the shuttle method [1, 2], which has been used by the heuristic arguments heretofore. Third, the dynamical output-feedback controller, using a practical robust control scheme, is devised that theoretically guarantees a bounded tracking error in the presence of matched disturbances and uncertainties: both the derivative of the drag acceleration tracking error and the matched disturbance are estimated using a second-order observer in the relative velocity domain; as a result, this dynamical output-feedback controller only needs the tracking error information, thus improving the robustness to navigation errors.

## II. Preliminary

Assuming the spherical and rotating Earth, the dimensionless equations of motion in the longitudinal plane for the RLV are listed below [20]:

$$\dot{r} = V \sin \gamma \tag{1}$$

$$\dot{V} = -D - \sin\gamma/r^2 + \Omega^2 r \cos\phi \left(\sin\gamma\cos\phi - \cos\gamma\sin\phi\cos\psi\right) \tag{2}$$

$$V\dot{\gamma} = L\cos\sigma + \left(V^2 - 1/r\right)\cos\gamma/r + 2\Omega V \cos\phi\sin\psi + \Omega^2 r \cos\phi \left(\cos\gamma\cos\phi + \sin\gamma\cos\psi\sin\phi\right) \tag{3}$$

The lift and the drag accelerations are defined by

$$L = \frac{R_e}{2m} \rho(r) V^2 S_{\text{ref}} C_L(r, M, \alpha) \tag{4}$$

$$D = \frac{R_e}{2m} \rho(r) V^2 S_{\text{ref}} C_D(r, M, \alpha) \tag{5}$$

The lift and drag coefficients $C_L$ and $C_D$ are functions of $r$, $M$, and $\alpha$, and, based on the kinetic theory of an isothermal gas in a uniform gravitational field, the well-known exponential approximation for the nominal atmospheric density $\hat{\rho}$ is defined [21]:

$$\hat{\rho} = \rho_0 e^{-\frac{R_e(r-1)}{H}} \tag{6}$$

Since the relative velocity is monotonically decreasing in the near-equilibrium glide phase, an equivalent model for Eqs. (1-3) can be derived by regarding the relative velocity as the independent variable of the dynamical system. Thus, the second-order velocity-varying/non-autonomous nonlinear system, ignoring the effects of the Earth's self-rotation, can be generally described by

$$x' = f(x, V, u) \tag{7}$$

We note that the subsequent analysis and design will be carried out in the velocity domain.

Suppose that a reference trajectory $\hat{x}$ together with the control input $\hat{u}$ is generated by a trajectory planning algorithm. The reference trajectory satisfies the dynamical constraint generally described by

$$\hat{x}' = \hat{f}(\hat{x}, V, \hat{u}) \tag{8}$$

where $\hat{f}(\hat{x}, V, \hat{u})$ is the nominal model used by the trajectory planning algorithm.

## III. Non-minimum Phase Property in Angle-of-Attack Modulation

In this section, we model a SISO plant with the unstable internal dynamics, describing the dynamical relationship between the drag acceleration and the angle-of-attack deviation, and then reveal the important control property that the input-output system of angle-of-attack modulation is the non-minimum phase system.

### A. Simplified Velocity-Varying Linear Model

By the first-order Taylor series expansion along the reference trajectory, the tracking error dynamics (referred to as *e*-dynamics henceforth) are given by

$$e' = Ae + B\delta u_1 + d^e \tag{9}$$

The matrix $A$ is the Jacobian matrix of the state variables:

$$A = \begin{bmatrix} a_{rr} & a_{r\gamma} \\ a_{\gamma r} & a_{\gamma\gamma} \end{bmatrix} \tag{10}$$

where

$$a_{rr} = -[V(\hat{D}\hat{r}^2 + \sin\hat{\gamma})^2]^{-1} V^2 \hat{r} \sin\hat{\gamma}(2\sin\hat{\gamma} - \hat{D}_r \hat{r}^3)$$

$$a_{r\gamma} = -[V(\hat{D}\hat{r}^2 + \sin\hat{\gamma})^2]^{-1} V^2 \hat{D}\hat{r}^4 \cos\hat{\gamma}$$

$$a_{\gamma r} = -[V(\hat{D}\hat{r}^2 + \sin\hat{\gamma})^2]^{-1} [\hat{L}_r \hat{r}^2 (\hat{D}\hat{r}^2 + \sin\hat{\gamma})\cos\hat{\sigma} - \hat{D}_r \hat{r}^2 (\hat{L}\hat{r}^2 \cos\hat{\sigma} + V^2 \hat{r}\cos\hat{\gamma} - \cos\hat{\gamma}) \\ - \hat{D}V^2 \hat{r}^2 \cos\hat{\gamma} + 2\hat{D}\hat{r}\cos\hat{\gamma} + 2\hat{L}\hat{r}\sin\hat{\gamma}\cos\hat{\sigma} + V^2 \cos\hat{\gamma}\sin\hat{\gamma}]$$

$$a_{\gamma\gamma} = -[V(\hat{D}\hat{r}^2 + \sin\hat{\gamma})^2]^{-1} [(V^2\hat{r} - 1)(-\hat{D}\hat{r}^2 \sin\hat{\gamma} - 1) - \hat{L}\hat{r}^2 \cos\hat{\sigma}\cos\hat{\gamma}]$$

The matrix $B$ is the Jacobian matrix of the angle-of-attack deviation:

$$B = \begin{bmatrix} b_{r\alpha} \\ b_{\gamma\alpha} \end{bmatrix} \tag{11}$$

where

$$b_{r\alpha} = [V(\hat{D}\hat{r}^2 + \sin\hat{\gamma})^2]^{-1} \hat{r}^4 V^2 \sin\hat{\gamma} \hat{D}_\alpha$$

$$b_{\gamma\alpha} = -[V(\hat{D}\hat{r}^2 + \sin\hat{\gamma})^2]^{-1}\hat{r}^2\{\hat{L}_\alpha \cos\hat{\sigma}(\hat{D}\hat{r}^2 + \sin\hat{\gamma}) - \hat{D}_\alpha[\hat{L}\hat{r}^2 \cos\hat{\sigma} + (V^2\hat{r} - 1)\cos\hat{\gamma}]\}$$

The linearization error is non-vanishing and can be described by

$$d^e = f(x, V, u) - \hat{f}(\hat{x}, V, \hat{u}) - Ae - B\delta u_1 \qquad (12)$$

The nominal model (8) can be approximated by certain algebraic relationships for the near-equilibrium glide trajectory [20]. Therefore, we attempt to simplify the matrices *A* and *B* by taking advantage of such algebraic relationships, so as to facilitate the following analysis and development of angle-of-attack modulation. On the basis of the near-equilibrium glide trajectory, four assumptions are established:

*Assumption 1:* The normalized radius $\hat{r}$ is approximated to unity.

*Assumption 2:* Except the term $-V^2 \sin\hat{\gamma}\hat{D}_r$ and the term $-V^2 \sin\hat{\gamma}\hat{D}_\alpha$, the remaining terms containing $\sin\hat{\gamma}$ are neglected. The term $\cos\hat{\gamma}$ is approximated to unity.

*Assumption 3:* Suppose that the radius has little effect on the aerodynamic coefficients. The involved term satisfies the below equality:

$$\hat{L}_r\hat{D}\cos\hat{\sigma} - \hat{D}_r\hat{L}\cos\hat{\sigma} = 0 \qquad (13)$$

*Assumption 4:* Based on the exponential atmosphere-density model (6), the term $\hat{D}(V^2 - 2)$ is approximately zero.

The four assumptions are sufficient conditions to simplify the original matrices *A* and *B*. From the perspective of flight mechanics of the RLV, it is easy to meet these assumptions. Based on Assumptions 1 through 4, the original matrices *A* and *B* can be reduced to

$$A = \begin{bmatrix} \bar{a}_{rr} & \bar{a}_{r\gamma} \\ \bar{a}_{\gamma r} & \bar{a}_{\gamma\gamma} \end{bmatrix} = \begin{bmatrix} V\hat{D}_r \sin\hat{\gamma}/\hat{D}^2 & -V/\hat{D} \\ \hat{D}_r(V^2-1)/V\hat{D}^2 & (V^2-1)+\hat{L}\cos\hat{\sigma}/V\hat{D}^2 \end{bmatrix} \qquad (14)$$

$$B = \begin{bmatrix} \bar{b}_{r\alpha} \\ \bar{b}_{\gamma\alpha} \end{bmatrix} = \begin{bmatrix} V^2 \sin\hat{\gamma}(\hat{C}_D)_\alpha/V\hat{D}\hat{C}_D \\ -\hat{L}_\alpha \cos\hat{\sigma}/V\hat{D} + [\hat{L}\cos\hat{\sigma} + (V^2-1)](\hat{C}_D)_\alpha/V\hat{D}\hat{C}_D \end{bmatrix} \qquad (15)$$

## B. Extended SISO System

Next, we are going to develop the SISO system of which input is the second-order derivative of the angle-of-attack deviation and output is the drag acceleration tracking error. Take the drag acceleration tracking error $y = D - \hat{D}$ as the output of interest; the output equation of the *e*-dynamics system can be put in the velocity-varying linear form:

$$y = Ce + \chi \delta u_1 + d^y \tag{16}$$

where the term $d^y = D - \hat{D} - \hat{D}_r \delta x_1 - \hat{D}_\alpha \delta u_1$ is the total modeling error of the linear form.

Given the static input-output relationship (16) (i.e., the relative degree of the SISO system is zero), the dynamic extension technique is adopted to obtain a dynamical input-output relationship with well-defined relative degree; that is, augment auxiliary state variable $\delta u_1'$, and then set state variable $\delta u_1$ equal to the output of a double-integrator system driven by virtual control input $\delta u_1^v$. In this way, the control input is delayed to a higher derivative of output. Besides, to simplify the implementation of the dynamic extension technique, we do not explicitly take the derivatives of the velocity-varying terms $A$, $B$, $C$ and $\chi$, while figure their effects as a modeling error part. Such a modeling scheme can be defended by the heuristic argument that the augmented dynamics are in a fast time-scale, whereas the velocity-varying elements in a slow time-scale.

As a result, the original second-order system is dynamically extended to the fourth-order SISO system represented in the standard input-affine form:

$$z' = f_z(z, V) + g \delta u_1^v + d^z \tag{17}$$

where

$$z = \begin{bmatrix} y & y' & \delta u_1 & \delta u_1' \end{bmatrix}^T$$

$$f_z(z,V) = \begin{bmatrix} z_2 & CA^2 \begin{bmatrix} C \\ CA \end{bmatrix}^{-1} \begin{bmatrix} z_1 \\ z_2 \end{bmatrix} + \left( CAB - CA^2 \begin{bmatrix} C \\ CA \end{bmatrix}^{-1} \begin{bmatrix} \chi \\ CB \end{bmatrix} \right) z_3 + \left( CB - CA^2 \begin{bmatrix} C \\ CA \end{bmatrix}^{-1} \begin{bmatrix} 0 \\ \chi \end{bmatrix} \right) z_4 & z_4 & 0 \end{bmatrix}^T$$

$$g = \begin{bmatrix} 0 & \chi & 0 & 1 \end{bmatrix}^T$$

$$d^z = \begin{bmatrix} 0 & d^Z & 0 & 0 \end{bmatrix}^T$$

where

$$d^Z = CAd^e + (CA)' e + (CB)' \delta u_1 + \chi' \delta u_1' + \left[ C'e + \chi' \delta u_1 + (d^y)' + Cd^e \right]' - CA^2 \begin{bmatrix} C \\ CA \end{bmatrix}^{-1} \begin{bmatrix} d^y \\ C'e + \chi' \delta u_1 + (d^y)' + Cd^e \end{bmatrix}$$

is the modeling error.

At this point, we obtain the more tractable fourth-order system with well-defined relative degree 2, and refer to this extended SISO system as the *z*-system henceforth. The issue of angle-of-attack modulation can be formulated as the robust output-feedback stabilization problem: design an output-feedback controller to stabilize the *z*-system in the presence of the modeling error $d^z$.

### C. SISO System in the Normal Form

Before designing the output-feedback controller using the feedback linearization tool, we need to put the *z*-system in the normal form, including second-order external dynamics and second-order internal dynamics. It is worth mentioning that we can only obtain the normal-form representation for the nominal part of the plant due to the disturbance $d^z$.

Consider a diffeomorphism $T(z) = \begin{bmatrix} \xi & \eta \end{bmatrix}^T$, the change of coordinates that is used to transform the coordinates into $(\xi \quad \eta)^T$, where $\xi = \begin{bmatrix} \xi_1 & \xi_2 \end{bmatrix}^T = \begin{bmatrix} z_1 & z_2 \end{bmatrix}^T$ is the external state vector, and $\eta = \begin{bmatrix} \eta_1 & \eta_2 \end{bmatrix}^T = \begin{bmatrix} N_1(z) & N_2(z) \end{bmatrix}^T$ is the external state vector such that

$$\eta = \begin{bmatrix} N_1(z) \\ N_2(z) \end{bmatrix} = \begin{bmatrix} z_3 \\ z_4 - \frac{1}{\chi} z_2 \end{bmatrix} \tag{18}$$

In the $(\boldsymbol{\xi} \ \boldsymbol{\eta})^T$-coordinate, the normal form of the $z$-system can be written as

$$\xi_1' = \xi_2 \tag{19}$$

$$\xi_2' = CA^2 \begin{bmatrix} C \\ CA \end{bmatrix}^{-1} \boldsymbol{\xi} + \left(CAB - CA^2 \begin{bmatrix} C \\ CA \end{bmatrix}^{-1} \begin{bmatrix} \chi \\ CB \end{bmatrix}\right) \eta_1 + \left(CB - CA^2 \begin{bmatrix} C \\ CA \end{bmatrix}^{-1} \begin{bmatrix} 0 \\ \chi \end{bmatrix}\right)\left(\eta_2 + \frac{1}{\chi}\xi_2\right) \tag{20}$$
$$+ \chi \delta u_1^v + d^Z$$

$$\eta_1' = \eta_2 + \frac{1}{\chi}\xi_2 \tag{21}$$

$$\eta_2' = -\frac{1}{\chi}\left[CA^2 \begin{bmatrix} C \\ CA \end{bmatrix}^{-1} \boldsymbol{\xi} + \left(CAB - CA^2 \begin{bmatrix} C \\ CA \end{bmatrix}^{-1} \begin{bmatrix} \chi \\ CB \end{bmatrix}\right) \eta_1 + \left(CB - CA^2 \begin{bmatrix} C \\ CA \end{bmatrix}^{-1} \begin{bmatrix} 0 \\ \chi \end{bmatrix}\right)\left(\eta_2 + \frac{1}{\chi}\xi_2\right)\right] \tag{22}$$
$$+ \frac{\chi'}{\chi^2}\xi_2 - \frac{1}{\chi}d^z$$

Based on the above SISO system, the stabilization problem in question can be solved by devising an input-output feedback linearization controller for the $\xi$-subsystem, provided that the $\eta$-subsystem is stable [13, 14]. Therefore, we need to check the stability of the $\eta$-subsystem.

**D. Non-minimum Phase System**

To determine the stability of the internal dynamics, an effective way is to examine the stability of the so called zero-dynamics [13, 14]. Substituting $A$, $B$, $C$, and $\chi$ in Eq. (20), and by algebraic manipulations, we obtain the zero-dynamics of the SISO system (19-22), given by

$$\eta_1' = \eta_2 \tag{23}$$

$$\eta_2' = \Gamma_1 \eta_1 + \Gamma_2 \eta_2 - \frac{1}{\hat{D}_\alpha} d^z \tag{24}$$

where the velocity-varying coefficients

$$\Gamma_1 = \frac{R_0}{H} \frac{\hat{L}}{\hat{D}} \frac{\cos \hat{\sigma}}{\hat{D}_\alpha} \left[\frac{(\hat{C}_L)_\alpha}{\hat{C}_L} - \frac{(\hat{C}_D)_\alpha}{\hat{C}_D}\right], \text{ and } \Gamma_2 = \frac{\left(V^2-1\right) + \hat{L}\cos\hat{\sigma}}{V\hat{D}^2} \tag{25}$$

It is worth pointing out that, owing to the simplified matrices $A$ and $B$, we can derive such concise yet accurate expressions of the coefficients $\Gamma_1$ and $\Gamma_2$, and identify two critical terms, $(\hat{C}_L)_\alpha/\hat{C}_L - (\hat{C}_D)_\alpha/\hat{C}_D$ and $(V^2-1)+\hat{L}\cos\hat{\sigma}$, closely related to the stability of the zero dynamics. Specifically, the term $(\hat{C}_L)_\alpha/\hat{C}_L - (\hat{C}_D)_\alpha/\hat{C}_D$ is determined by the reference angle-of-attack profile and is slowly velocity-varying in accordance with the hypersonic aerodynamics. The term $(V^2-1)+\hat{L}\cos\hat{\sigma}$ is determined by the rate of the flight-path angle in the near-equilibrium glide phase and is also slowly velocity-varying.

Because of the slowly velocity-varying properties of $\Gamma_1$, $\Gamma_2$, and $d^z$, together with the short-period characteristics of the zero-dynamics, it is practical to assume that on a small time horizon the coefficients $\Gamma_1$ and $\Gamma_2$ as well as the disturbance $d^z$ are velocity-invariant. Based on this assumption, two flight conditions rendering the zero-dynamics unstable are derived and are listed below:

*Flight Condition 1*: $\Gamma_1 < 0$, and $\Gamma_2 < 0$. In this case, the zero-dynamics have a couple of complex eigenvalues in the left-half complex plane in the velocity domain, and Flight Condition 1 can be reduced to the inequalities:

$$\frac{\partial \hat{C}_L}{\partial \hat{C}_D} - \frac{\hat{C}_L}{\hat{C}_D} < 0 \text{ and } (V^2-1)+\hat{L}\cos\hat{\sigma} < 0 \qquad (26)$$

where $\partial \hat{C}_L/\partial \hat{C}_D$ is the slope of the drag polar of the reference trajectory, and $\hat{C}_L/\hat{C}_D$ is the lift-to-drag ratio of the reference trajectory. The two inequalities have explicitly physical meaning: the zero-dynamics have unstable when the RLV flies on the "back side" of the lift-to-drag vs. angle-of-attack curve (i.e., $\partial \hat{C}_L/\partial \hat{C}_D - \hat{C}_L/\hat{C}_D < 0$) on the condition that the reference rate of the flight-path angle is negative (i.e., $(V^2-1)+\hat{L}\cos\hat{\sigma} < 0$).

*Flight Condition 2*: $\Gamma_1 > 0$. In this case, the zero-dynamics have two real eigenvalues: a positive one and a negative one, and Flight Condition 2 can be reduced to the inequality:

$$\frac{\partial \hat{C}_L}{\partial \hat{C}_D} - \frac{\hat{C}_L}{\hat{C}_D} > 0 \qquad (27)$$

which likewise implies that the zero-dynamics are unstable when the RLV operates on the "front side" of the lift-to-drag vs. angle-of-attack curve.

To the best of the authors' knowledge, this is the first time in the area of entry guidance to reveal the non-minimum phase property for angle-of-attack modulation. The SISO system described by the velocity-varying system (19-22) is the non-minimum phase system under one of the two flight conditions. The internal dynamics will diverge if no controller is designed to actively stabilize the internal dynamics. Since the internal dynamics are composed of the augmented dynamics, $\delta u_1$ and $\delta u_1'$, the divergence implies that the angle-of-attack deviation diverges and has the great potential risk to violate the prescribed working range constraint, mainly determined by the TPS, the dynamic stability of the RLV, the ACS, and the cross-range requirements. In the design of the output-feedback controller, therefore, it is required to actively stabilize the internal dynamics, correspondingly driving the actual angle-of-attack back to its reference profile.

## IV. Angle-of-Attack Modulation

In this section, an output-feedback controller is designed to robustly stabilize the non-minimum phase system based on the cooperative control strategy depicted in Fig. 1. As shown in Fig. 1, bank-angle modulation $\Phi(x,V,\hat{D})$ is the primary control means to approach the trajectory control problems on a long-period basis; angle-of-attack modulation $N(\xi_1,V)$ is employed to achieve guidance tasks in a short-period basis when the RLV reverses the bank angle or encounters fast time-varying disturbances; once the bank reversal is completed, the control $\delta u_2 = K_\alpha(\eta)$ starts to work to actively stabilize the unstable internal dynamics.

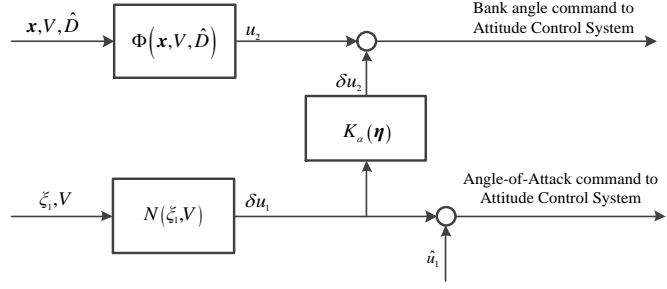

**Fig. 1  Schematic diagram of the cooperative control strategy.**

## A. Dynamical Output-Feedback Controller

In this subsection, we devise the controller for angle-of-attack modulation:

$$\delta u_1 = N(\xi_1, V) = \iint L(\xi_1, \delta u_1, \delta u_1', V) dV \tag{28}$$

where $\delta u_1^v = L(\xi_1, \delta u_1, \delta u_1', V)$ is the dynamical output-feedback controller for the virtual control input.

To design the dynamical output-feedback controller, the disturbance observer is incorporated into the feedback linearization controller, and the structure of thus combined controller is illustrated in Fig. 2—the disturbance observer attempts to estimate the modeling error $d^z$, whose estimated value is used to generate the corrective input to approximately eliminate the adverse influences on the tracking errors.

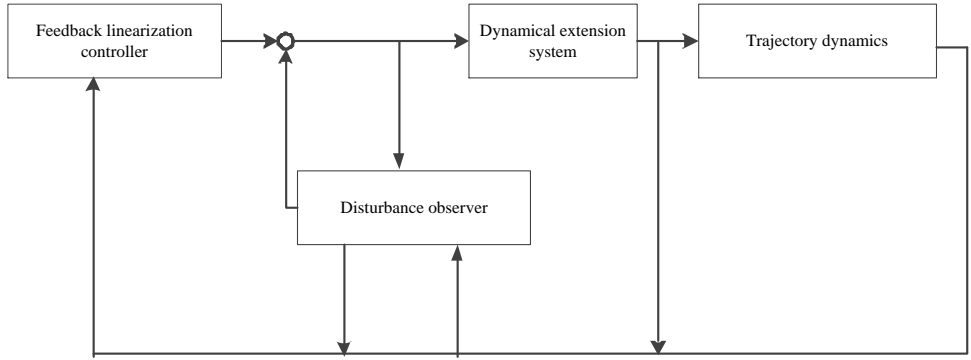

**Fig. 2  Control structure of the closed system.**

Accordingly, the dynamical output-feedback controller includes the feedback linearization controller $u_{FL}$ and the disturbance observer $u_{DC}$:

$$\delta u_1^v = L(\xi_1, \delta u_1, \delta u_1', V) = u_{FL} + u_{DC} \tag{29}$$

The feedback linearization controller $u_{FL}$ consists of two parts:

$$u_{FL} = \frac{1}{\chi}\left[\left(\delta u_1^v\right)_m + \Lambda\right] \tag{30}$$

where the term $\left(\delta u_1^v\right)_m$ is used to cancel the nominal plant in the external dynamics, taking the form:

$$\left(\delta u_1^v\right)_m = -\left[\bar{a}_{rr}\xi_2 + \left(\bar{a}_{r\gamma}\bar{a}_{\gamma r} - \bar{a}_{rr}\bar{a}_{\gamma\gamma}\right)\xi_1 + \bar{a}_{r\gamma}\left(\bar{c}_r\bar{b}_{\gamma\alpha} - \bar{a}_{\gamma r}\chi\right)\eta_1 - \bar{a}_{\gamma\gamma}\chi\eta_2\right] \tag{31}$$

The term $\Lambda$ is employed to specify the nominal transient performance of the tracking errors, given by

$$\Lambda = -2\varsigma\omega_n\xi_2 - \omega_n^2\xi_1 \tag{32}$$

where $\varsigma$ is a damping ratio and $\omega_n$ is nature frequency in the velocity domain. The transient performance can be shaped by adjusting $\varsigma$ and $\omega_n$, so as to meet certain engineering requirements. In addition, the tuning of natural frequency $\omega_n$ needs to guarantee the influences of angle-of-attack modulation in a short-period timescale.

The scheme of the disturbance observer-based control techniques is to make the resulting closed system behave as the nominal plant through dynamically compensating the real system by feeding the estimated disturbances back to the input. Thus, the disturbance observer needs to meet the following ideal relation:

$$u_{DC} = -\frac{1}{\chi}d^z \tag{33}$$

where the matched disturbance $d^z$ can be taken as

$$d^z = -\left[\bar{a}_{rr}\xi_2 + \left(\bar{a}_{r\gamma}\bar{a}_{\gamma r} - \bar{a}_{rr}\bar{a}_{\gamma\gamma}\right)\xi_1 + \bar{a}_{r\gamma}\left(\bar{c}_r\bar{b}_{\gamma\alpha} - \bar{a}_{\gamma r}d\right)\eta_1 - \bar{a}_{\gamma\gamma}\chi\eta_2 + \chi\left(\delta u_1^v\right) - \xi_2'\right] \tag{34}$$

This disturbance observer cannot be directly implemented because the variables $\xi_2$ and $\xi_2'$ are not available. Therefore, we employ a second-order observer to estimate the state variable $\xi_2$ and the disturbance $d^z$. The observer is in the below form:

$$\left(\hat{\xi}_2\right)' = \bar{a}_{rr}\hat{\xi}_2 + \hat{d}^z + \tau_1\left(\xi_2 - \hat{\xi}_2\right) + \chi\left(\delta u_1^v\right) + \left(\bar{a}_{r\gamma}\bar{a}_{\gamma r} - \bar{a}_{rr}\bar{a}_{\gamma\gamma}\right)\xi_1 + \bar{a}_{r\gamma}\left(\bar{c}_r\bar{b}_{\gamma\alpha} - \bar{a}_{\gamma r}\chi\right)\eta_1 - \bar{a}_{\gamma\gamma}\chi\hat{\eta}_2 \tag{35}$$

$$\left(\hat{d}^z\right)' = \tau_2\left(\xi_2 - \hat{\xi}_2\right) \tag{36}$$

where $\hat{\xi}_2$ is the estimated value of $\xi_2$, $\hat{\eta}_2$ is the estimated value of $\eta_2$ such that $\hat{\eta}_2 = \delta u_1' - \frac{1}{\chi} \hat{\xi}_2$, $\hat{d}^z$ is the estimated value of $d^z$, and $\tau_1$ and $\tau_2$ are the observer gains. The estimation error vector

$$\begin{bmatrix} \tilde{\xi}_2 \\ \tilde{d}^z \end{bmatrix} = \begin{bmatrix} \xi_2 - \hat{\xi}_2 \\ d^z - \hat{d}^z \end{bmatrix} \tag{37}$$

satisfies the dynamical equations:

$$\tilde{\xi}_2' = \left( \bar{a}_{rr} + \bar{a}_{\gamma\gamma} - \tau_1 \right) \tilde{\xi}_2 + \tilde{d}^z \tag{38}$$

$$\left( \tilde{d}^z \right)' = -\tau_2 \tilde{\xi}_2 \tag{39}$$

The tuning of $\tau_1$ and $\tau_2$ ensures that $\lim_{V \to \infty} \tilde{\xi}_2 = 0$ and $\lim_{V \to \infty} \tilde{d}^z = 0$. We can specify the convergence rate of the observer by appropriately choosing these two gains.

To avoid amplifying the measurement noise, it is suggested to eliminate the calculation of $\xi_1'$. Thus, we replace $\hat{\xi}_2$ by $\bar{\xi}_2 + \tau_1 \xi_1$ and $\hat{d}^z$ by $\bar{d}^z + \tau_2 \xi_1$. Correspondingly, the above observer equations can be taken as

$$\left( \bar{\xi}_2 \right)' = \left( \bar{a}_{rr} + \bar{a}_{\gamma\gamma} - \tau_1 \right) \bar{\xi}_2 + \bar{d}^z + \chi \left( \delta u_1^v \right) + \left( \bar{a}_{r\gamma} \bar{a}_{\gamma r} - \bar{a}_{rr} \bar{a}_{\gamma\gamma} + \bar{a}_{rr} \tau_1 + \bar{a}_{\gamma\gamma} \tau_1 - \tau_1^2 + \tau_2 \right) \xi_1 \\ + \bar{a}_{r\gamma} \left( \bar{c}_r \bar{b}_{\gamma\alpha} - \bar{a}_{\gamma r} \chi \right) \eta_1 - \bar{a}_{\gamma\gamma} \chi \delta u_1' \tag{40}$$

$$\left( \bar{d}^z \right)' = -\tau_2 \bar{\xi}_2 - \tau_1 \tau_2 \xi_1 \tag{41}$$

Replacing $d^z$ by its estimate $\hat{d}^z$, and replacing $\xi_2$ by its estimate $\hat{\xi}_2$ in Eq. (29), we arrive at the robustly dynamical output-feedback controller.

### B. Stabilization of the Internal Dynamics

In this subsection, we realize the stabilization of the internal dynamics using the internal-state feedback control scheme. Based on the cooperative control strategy, the internal-state feedback controller is employed for bank-angle modulation:

$$|\delta u_2| = K_\alpha \left( \eta_1, \hat{\eta}_2 \right) = k_1 \eta_1 + k_2 \hat{\eta}_2 = k_1 \eta_1 + k_2 \left( \delta u_1' - \frac{1}{\chi} \hat{\xi}_2 \right) \tag{42}$$

where $k_1$ and $k_2$ are the feedback gains.

Compared to the stabilization scheme used in the shuttle entry guidance [1], the major contribution here is that our method is based on the rigorous control mechanism. Concerning the internal dynamics in question, $\xi$ can be regarded as the control input; $\eta$ can be regarded as the state. In this context, the bank angle actually makes the output track the expected external dynamics that are used to stabilize the internal dynamics. In other words, instead of eliminating the tracking errors, the bank angle actively introduces the tracking errors to stabilize the internal dynamics. Although this controller can effectively stabilize the unstable internal dynamics, it has two weaknesses. First, since the control means is bank-angle modulation, this controller does not work in some situations. As a result, the angle-of-attack deviation diverges on a short time interval. Second, there is a transient-tracking-accuracy limit because this controller needs to actively introduce the tracking errors to stabilize the unstable internal dynamics.

## V. Numerical Simulations

### A. RLV Model

In the following numerical simulations, the RLV data are used. The nominal aerodynamics coefficients are approximated by fitting the data in the hypersonic regime:

$$\hat{C}_L = 0.12457 - 0.02437\alpha + 0.00309\alpha^2 - 3.66023 \times 10^{-5} \alpha^3 \qquad (43)$$

$$\hat{C}_D = 0.32083 - 0.02850\alpha + 0.00155\alpha^2 - 9.42499 \times 10^{-7} \alpha^3 \qquad (44)$$

where $\alpha$ is in degrees, and the reference area is 5 m$^2$.

The nominal angle-of-attack reference profile is designed:

$$\hat{\alpha} = \begin{cases} 40 \text{ deg} & M \geq 12 \\ -4.3333 + 7.3611M - 0.3056M^2 \text{ deg} & 3 \leq M \leq 12 \end{cases} \qquad (45)$$

In addition, second-order models are used to emulate the dynamical response of the ACS. Take the bank-angle as an example; the second-order system of interest is given below:

$$\ddot{\sigma} = -2\xi_\sigma \omega_\sigma \dot{\sigma} - \omega_\sigma^2 \sigma + \omega_\sigma^2 \sigma_{com} \qquad (46)$$

where $\sigma_{com}$ is the command computed using the controller for bank-angle-modulation, $\xi_\sigma$ and $\omega_\sigma$ are two parameters that represent the lateral dynamical characteristics of the ACS, and $\sigma$, $\dot{\sigma}$, and $\ddot{\sigma}$ are the actual bank angle, the first-order derivative of the bank angle with respect to time, and the second-order derivative of the bank angle with respect to time, respectively. The actual bank angle is obtained via an integration logic that is developed to enforce the hard constraints: $\sigma_{max} = 80 \text{ deg}$, $\dot{\sigma}_{max} = 5 \text{ deg/sec}$, and $\ddot{\sigma}_{max} = 1.7 \text{ deg/sec}^2$.

**B. Comparison Results**

In this subsection, compared to the shuttle method [1, 2], nominal and dispersion tests are conducted to demonstrate the performance of the dynamical output-feedback controller. In the following numerical tests, the reference trajectory consisting of a constant heating-rate segment and a pseudo-equilibrium gliding segment is generated using a trajectory planning algorithm [22]. For the sake of comparisons, both the dynamical output-feedback controller and the shuttle method use the same bank angle controller. Besides, angle-of-attack modulation starts to work at a velocity of 7200 m/s where the RLV has passed the high heating region. A stressful case is configured for the particular performance of restraining the oscillations induced by the bank reversals: the six bank reversals are arranged that in sequence occur at 7000 m/s, 6000 m/s, 5000 m/s, 4000 m/s, 3000 m/s, and 2000 m/s. In fact, a nominal RLV entry mission has no more than four bank reversals, but a large number of the bank reversals may be demanded to meet severe requirements of potential RLV missions, such as overflight considerations and landing site online update. In this respect, it is useful to enhance the tracking performances in the presence of a large number of the bank reversals.

*1. Nominal Test*

Figures 3 shows that the actual drag acceleration profiles of the two methods are nearly identical to the reference profile, and illustrates that both the two methods can achieve the satisfactory tracking performance in the nominal conditions. To closely examine the tracking errors, given in Fig. 4, we can observe that, after the first peaking (caused by the limited respond rate of bank-angle modulation), the dynamical output-feedback controller produces a tracking error on the order of $10^{-3}$ $g$ in magnitude, while the shuttle method roughly has an error on the order of $10^{-2}$ $g$ in magnitude. Besides, it is evident that the tracking errors of the two methods are practically identical when the velocity is bigger than 6000 m/s, and the differences in the tracking errors become larger as the velocity decreases. Such differences are mainly caused by the influences of the trajectory dynamics, and demonstrate that the tracking error in the high dynamic pressure regime indeed can be reduced by explicitly dealing with the influences of the trajectory dynamics of interest. The extent of the influences of the trajectory dynamics on the drag acceleration largely depends on the dynamic pressure: when the RLV flies in the low dynamic pressure regime, the influences can be ignored, and, as the dynamic pressure increases on entering denser atmosphere, the influences becomes more and more obvious. Therefore, thanks to considering the influences of the trajectory dynamics, the dynamical output-feedback controller can enhance the tracking performance, especially in the high dynamic pressure regime.

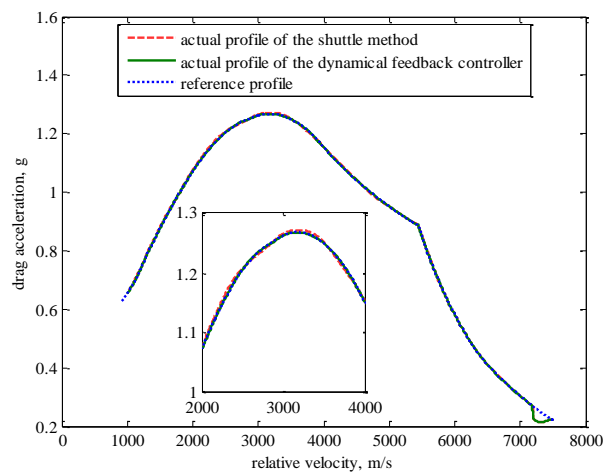

**Fig. 3   Drag acceleration profiles.**

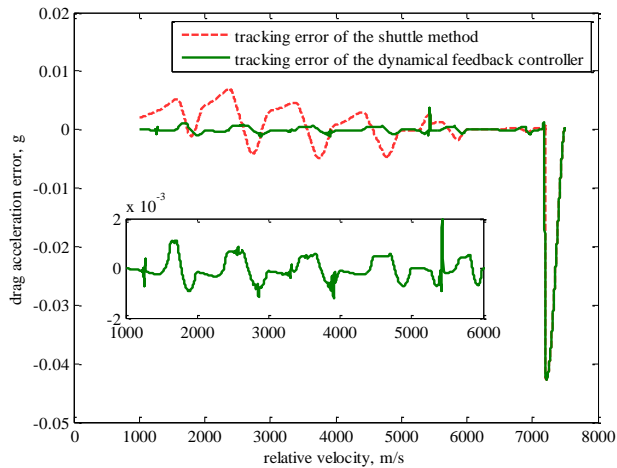

**Fig. 4   Drag acceleration tracking errors.**

As shown in Fig. 5, there exist seven spikes in the angle-of-attack deviation profiles. The first spike in the right side of the figure is used to eliminate the peaking error caused by the capability limitation of bank-angle modulation, and the remaining six spikes are employed to restrain the adverse effects of the phugoid motions induced by the intended six bank reversals. The trough around 5000 m/s is used to eliminate the adverse effects of the segment transition. With respect to each of these seven spikes, the angle-of-attack deviation continues to diverge until the bank reversal is completed due to the non-minimum phase property. Once this happens, the internal-dynamics stabilization controller takes effect to drive the deviation back to zero.

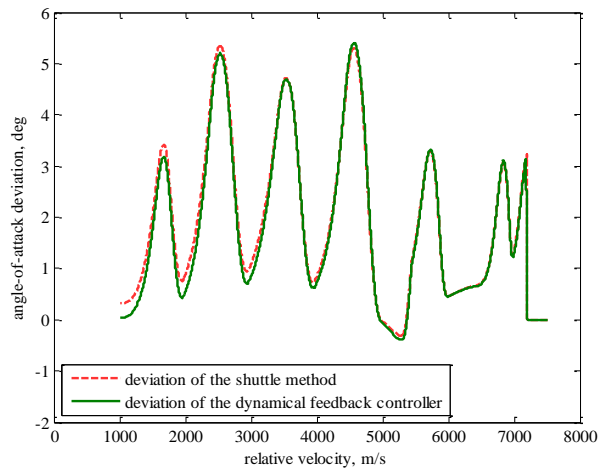

**Fig. 5   Angle-of-attack deviations.**

It is shown in Fig. 6 that the differences of the flight-path angle profiles can be negligible. The trajectory dynamics are represented by the up-and-down motions in the flight-path angle profiles. Figure 7 presents that the two bank angle profiles are practically identical. As shown in Fig. 8, the peaking in the estimated derivative of the tracking error is caused by the high observer gains. The adverse influences of this peaking are restrained because of the effective smoothing function of the augmented double-integrator (i.e., the extended dynamical system). Additionally, as the relative velocity decreases, the magnitude of the estimated disturbances is increasing. This phenomenon demonstrates the approximation precision of the simplified velocity-varying linear model descends. This approximation error is mainly caused by the constantly decreasing the flight-path angle, slightly violating Assumption 2 used in the derivation. Owing to the disturbance observer, the tracking errors can be rendered within the small range.

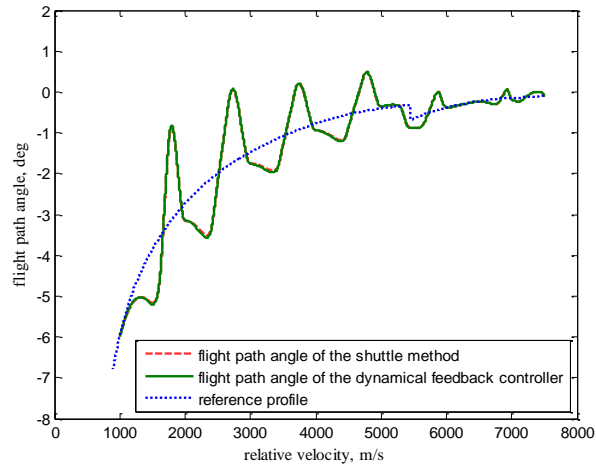

**Fig. 6  Flight-path angle profiles.**

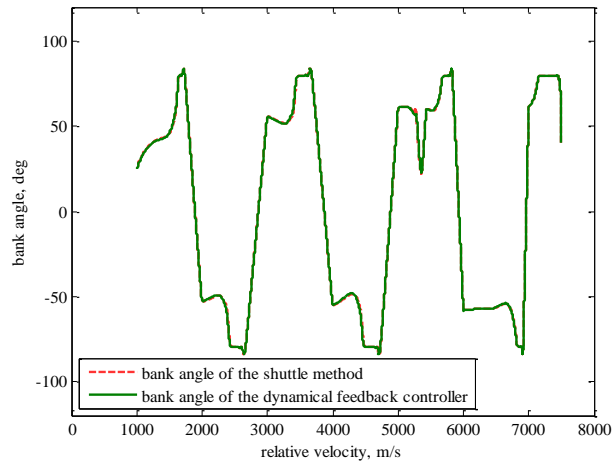

**Fig. 7    Bank angle profiles.**

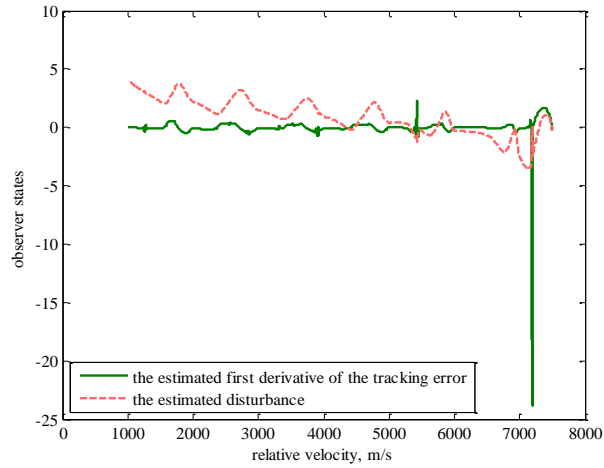

**Fig. 8    Disturbance Observer States.**

*2. Dispersion Test*

In order to assess the robustness of the dynamical output-feedback controller to environmental dispersions and aerodynamic uncertainties, a 500-sample Monte Carlo dispersion analysis is conducted. In this test, both the parameters of the feedback linearization controller and the observer gains are identical to the ones used in the nominal test. Table 1 summarizes the input data that are currently dispersed. The dispersed atmosphere model is developed to provide various dispersion parameters of the atmosphere. Besides, a constant error of the inflight estimated angle-of-attack is added into the Monte Carlo dispersion analysis.

**Table 1 Monte Carlo dispersion levels**

| Parameter | Dispersion | Mean | 3-$\sigma$ or Min/Max |
|---|---|---|---|
| EI Flight-Path Angle | Gaussian | 0 | 0.1 deg |
| Aero Coefficient of Lift | Gaussian | Aero-data | 10 % |
| Aero Coefficient of Drag | Gaussian | Aero-data | 15 % |
| The Atmosphere | Dispersed model | 1976Std Atmosphere | Dispersed model |
| The Inflight Estimated Angle-of-Attack Error | Constant | 0.1 deg | 0.1 deg |

Compared Fig. 9 with Fig. 10, it is evident that the accuracy of the dynamical output-feedback controller is better than that of the shuttle method. The dispersion magnitude of the drag acceleration errors of the dynamical output-feedback controller is less than about $2\times 10^{-3}$ g when the relative velocity is below 7000 m/s. Such small tracking errors demonstrate that the disturbance observer timely estimates the disturbances and compensates the adverse influences on the output. As for the shuttle method, the dispersion magnitude of the errors is larger, and the magnitude of the errors is up to about $2\times 10^{-2}$ g. Besides, the Monte Carlo samples diverge from its nominal trajectory. It is caused by the added constant error of the inflight estimated angle-of-attack. In the light of the baseline used in command computation, the shuttle method uses the current angle-of-attack, whereas the dynamical output-feedback controller is based on the reference angle-of-attack. Due to the dependence on the actual angle-of-attack, the shuttle method is more sensitive to estimated errors of the navigation system－the estimated error magnitude can be enlarged in the presence of the wind. Figures 11 and 12 display the angle-of-attack profiles. Due to the large drag acceleration errors at the start of angle-of-attack modulation, the commanded angle-of-attack deviations are considerable. In fact, these large deviations could be effectively eliminated by incorporating a limiter of the angle-of-attack, but scarifying a certain amount of the tracking accuracy.

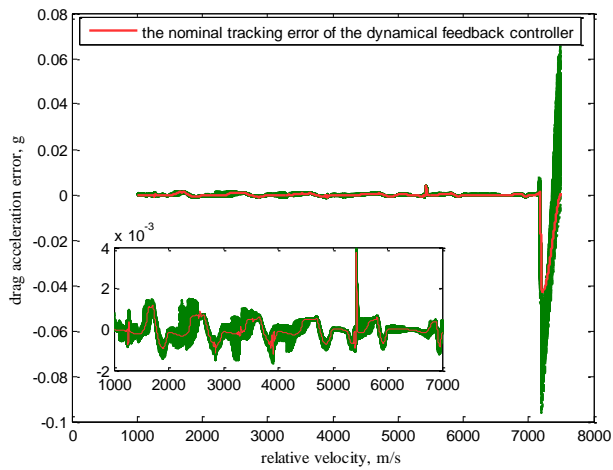

**Fig. 9    Drag acceleration tracking errors of the dynamical output-feedback controller.**

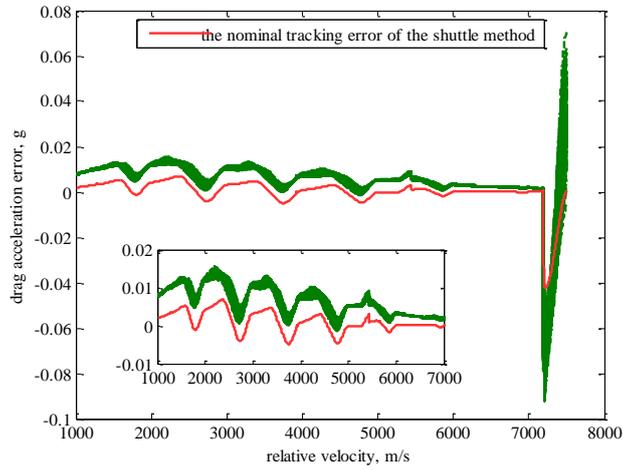

**Fig. 10    Drag acceleration tracking errors of the shuttle method.**

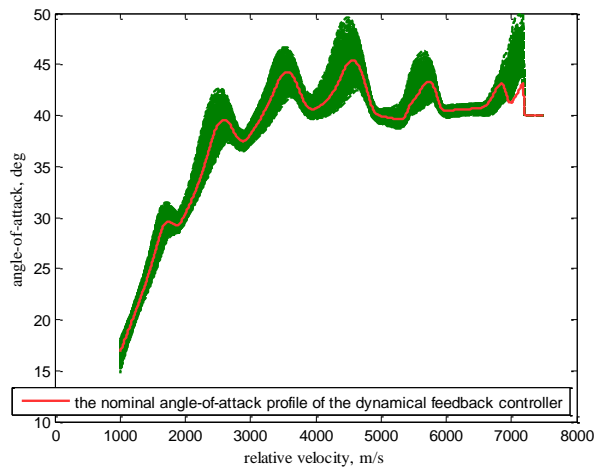

**Fig. 11    Angle-of-attack profiles of the dynamical output-feedback controller.**

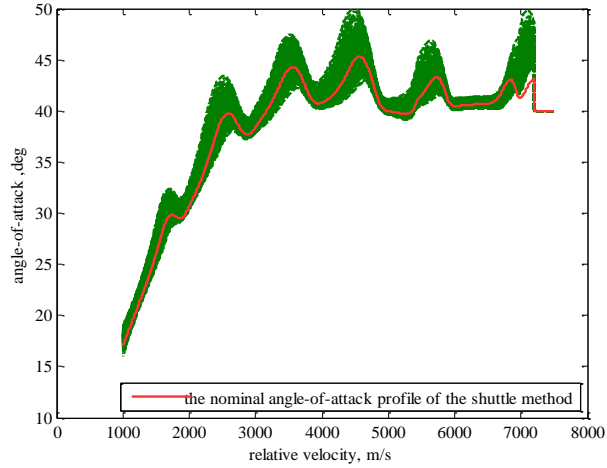

**Fig. 12 Angle-of-attack profiles of the shuttle method from Monte Carlo Analysis.**

## VI. Conclusion

The capability of angle-of-attack modulation can be integrated into the entry guidance methods to provide the supplementary trajectory control means during the bank reversals as well as to compensate for the short-period disturbances such as the atmospheric density gradients. This paper rigorously formulates the issue of angle-of-attack modulation as the robust output stabilization of the non-minimum phase system. By examining the stability of the zero dynamics of the extended fourth order SISO system, it is found that the input-output system for angle-of-attack modulation has the non-minimum phase property under two common flight conditions: the first is that the RLV flies on the "back side" of the lift-to-drag vs. angle-of-attack curve on the condition that the reference rate of the flight-path angle is negative; the second is that the RLV operates on the "front side" of the lift-to-drag vs. angle-of-attack curve. From the perspective of control theory, this finding provides a rigorous interpretation of the divergence characteristic of angle-of-attack modulation. Based on the cooperative control strategy, the dynamical output-feedback controller, together with the internal-states feedback controller, is devised to complete angle-of-attack modulation. The drag acceleration tracking performance of the dynamical output-feedback controller on the accuracy is practically identical to that of the existing shuttle method in the nominal conditions, except in the high dynamic pressure regime the dynamical output-feedback controller can enhance the performance because of considering the influences of the trajectory dynamics; besides, the dynamical output-feedback controller is more robust to the various uncertainties, disturbances and the navigation errors, in part due to the disturbance observer. The proposed method for angle-of-attack modulation can be applicable to the RLV with the attitude control capability of maintaining the required angle-of-attack, and can be readily incorporated into the available reference-profile tracking law with few adjustments, thus enhancing the transient guidance performance and the robustness of entry guidance.

**Reference**

[1] Harpold, J. C., and Graves, C. A. Jr., "Shuttle Entry Guidance," *Journal of the Astronautical Sciences*, Vol. 27, No. 3, 1979, pp. 239- 268.

[2] Harpold, J. C., and Gavert, D. E., "Space Shuttle Entry Guidance Performance Results," *Journal of Guidance*, Vol. 6, No. 6, 1983, pp. 442-447.

doi: 10.2514/3.8523

[3] Putnam, Z. R., Grant, M. J., Kelly, J. R., Braun, R. D., and Krevor, Z. C., "Feasibility of Guided Entry for a Crewed Lifting Body Without Angle-of-Attack Control," *Journal of Guidance, Control, and Dynamics*, Vol. 37, No. 3, 2014, pp. 729-740.

doi: 10.2514/1.62214

[4] Hanson, J. M., Coughlin, D. J., Dukeman, G. A., Mulqueen, J. A., and McCarter, J. W., "Ascent, Transition, Entry, and Abort Guidance Algorithm Design for the X-33 Vehicle," *Guidance, Navigation, and Control Conference and Exhibit*, AIAA Paper 1998-4409, Aug. 1998.

doi: 10.2514/6.1998-4409

[5] Hanson, J. M., and Jones, R. E., "Test Results for Entry Guidance Methods for Space Vehicles," *Journal of Guidance, Control, and Dynamics*, Vol. 27, No. 6, 2004, pp.960-966.

doi: 10.2514/1.10886

[6] Leavitt, J. A., and Mease, K. D., "Feasible Trajectory Generation for Atmospheric Entry Guidance," *Journal of Guidance, Control, and Dynamics*, Vol. 30, No. 2, 2007, pp. 473-481.

doi: 10.2514/1.23034

[7]  Mease, K. D., Chen, D. T., Teufel, P., and Schonenberger, H., "Reduced-Order Entry Trajectory Planning for Acceleration Guidance," *Journal of Guidance, Control, and Dynamics*, Vol. 25, No. 2, 2002, pp. 257-266.

doi: 10.2514/2.4906

[8]  Ping, L., "Regulation About Time-Varying Trajectories: Precision Entry Guidance Illustrated," *Journal of Guidance, Control, and Dynamics*, Vol. 22, No. 6, 1999, pp. 784-790.

doi: 10.2514/2.4479

[9]  Lu, P., "Entry Guidance: A Unified Method," *Journal of Guidance, Control, and Dynamics*, Vol. 37, No. 3, 2014, pp. 713-728.

doi: 10.2514/1.62605

[10]  Lu, P., "Entry Guidance Using Time-Scale Separation in Gliding Dynamics," *Journal of Spacecraft and Rockets*, Vol. 52, No. 4, 2015, pp. 1253-1258.

doi: 10.2514/1.A33295

[11]  McWhorter, L. B., and Reed, M, Space Shuttle Entry Digital Autopilot, NASA SP-2010-3408, 2010, pp. 79-81.

[12]  Masciarelli, J. P., Westhelle, C. H., and Graves, C. A., "AEROCAPTURE GUIDANCE PERFORMANCE FOR THE NEPTUNE ORBITER," *Atmospheric Flight Mechanics Conference and Exhibit*, AIAA Paper 2004-4954, Aug. 2004.

doi: 10.2514/6.2004-4954.

[13]  Isidori, A., *Nonlinear Control Systems*, 3rd ed., Springer-Verlag, London, 1995, chaps. 1, 5.

[14]  Isidori, A., "The zero dynamics of a nonlinear system: From the origin to the latest progresses of a long successful story," *European Journal of Control*, Vol. 19, May, 2013, pp. 369-378.

doi: 10.1016/j. ejcon.2013.05.014

[15]  Khalil, H. K., *Nonlinear Systems*, 3rd ed., Prentice Hall: Upper Saddle River, New Jersey, 2002, chap. 4.

[16] Isidori, A, "A Tool for Semiglobal Stabilization of Uncertain Non-Minimum-Phase Nonlinear Systems via Output Feedback," *IEEE Trans. Autom. Control*, Vol. 45, No. 10, 2000, pp.1817-1827.

doi: 10.1109/TAC.2000.880972

[17] Nazrulla, S., and Khalil, H. K., "Robust Stabilization of Non-Minimum Phase Nonlinear Systems Using Extended High-Gain Observers," *IEEE Trans. Autom. Control*, Vol. 56, No. 4, 2011, pp. 802-813.

doi: 10.1109/TAC.2010.2069612

[18] Freidovich, L. B., and Khalil, H. K., "Performance Recovery of Feedback-Linearization-Based Designs," *IEEE Trans. Autom. Control*, Vol. 53, No. 10, 2008, pp. 2324-2334.

doi: 10.1109/TAC.2008.2006821

[19] Chen, W. H., Ohnishi, K., and Guo, L., "Advances in Disturbance/Uncertainty Estimation and Attenuation," *IEEE Trans. Ind. Electron.*, Vol. 62, No. 9, 2015, pp. 5758-5762.

doi: 10.1109/TIE.2015.2453347

[20] Nguyen, X. V., Adolf, B., and Robert, D. C., *Hypersonic and Planetary Entry Flight Mechanics*, University of Michigan Press, Ann Arbor, 1980, pp. 19- 28.

[21] Chapman, D. R., "An Approximate Analytical Method for Studying Entry into Planetary Atmospheres," NACA TN-4276, May 1958.

[22] Li, H., Zhang, R., Li, Z., and Zhang, R., "New Method to Enforce Inequality Constraints of Entry Trajectories," *Journal of Guidance, Control, and Dynamics*, Vol. 35, No. 5, 2012, pp.1662-1667.

doi: 10.2514/1.56937